\documentclass[doublecol,amsmath,amssymb]{epl2}%showpacs,preprintnumbers,
\usepackage{amssymb}
\usepackage{graphicx}% Include figure files
\usepackage{dcolumn}% Align table columns on decimal point
\usepackage{bm}% bold math
\usepackage{hhline}
\usepackage{longtable}
\usepackage[usenames]{color}
\begin{document}

\title{Random walks on the Apollonian network with a single trap}
\shorttitle{Random walks on Apollonian network with a single trap}

\author{Zhongzhi Zhang\inst{1,2} \footnote{ \email{zhangzz@fudan.edu.cn} } \and Jihong Guan\inst{3} \footnote{\email{jhguan@tongji.edu.cn}} \and Wenlei Xie\inst{1,2}  \and Yi Qi \inst{1,2} \and  Shuigeng Zhou\inst{1,2} \footnote{\email{sgzhou@fudan.edu.cn}}}
\shortauthor{Zhongzhi Zhang, Jihong Guan, Wenlei Xie, Yi Qi,
Shuigeng Zhou}

 \institute{
  \inst{1} School of Computer Science, Fudan University, Shanghai 200433, China\\
  \inst{2} Shanghai Key Lab of Intelligent Information Processing, Fudan University, Shanghai 200433, China\\
 \inst{3} Department of Computer Science and Technology,
Tongji University, 4800 Cao'an Road, Shanghai 201804, China}

\date{\today}

\begin{abstract}{
Explicit determination of the mean first-passage time (MFPT) for
trapping problem on complex media is a theoretical challenge. In
this paper, we study random walks on the Apollonian network with a
trap fixed at a given hub node (i.e. node with the highest degree),
which are simultaneously scale-free and small-world. We obtain the
precise analytic expression for the MFPT that is confirmed by direct
numerical calculations. In the large system size limit, the MFPT
approximately grows as a power-law function of the number of nodes,
with the exponent much less than 1, which is significantly different
from the scaling for some regular networks or fractals, such as
regular lattices, Sierpinski fractals, T-graph, and complete graphs.
The Apollonian network is the most efficient configuration for
transport by diffusion among all previously studied structure.}
\end{abstract}

 %\pacs{89.75.Da}{Systems obeying scaling laws}

%89.20.Hh World Wide Web, Internet
%89.75.Da Systems obeying scaling laws
%89.75.Fb Structures and organization in complex systems
%89.75.Hc Networks and genealogical trees
%89.75.-k Complex systems
%05.10.-a Computational methods in statistical physics and nonlinear
%                dynamics
%05.45.Xt Synchronization; coupled oscillators
%{02.10.Ox}{Combinatorics; graph theory} \

 \maketitle

\section{Introduction}

Trapping is an integral major theme of random walks
(diffusion)~\cite{HaBe87,MeKl04,BuCa05}, which is relevant to a wide
range of applications and has led to a growing number of theoretical
and practical investigation over the past
decades~\cite{SoMaBl97,No77,LlMa01,FoPiReSa07,CoBeTeVoKl07,CoTeVoBeKl08}.
The trapping problem, first introduced in~\cite{Mo69}, is in fact a
random-walk issue, where a trap is positioned at a given location,
which absorbs all particles visiting it. The primarily interesting
quantity closely related to trapping problem is the average trapping
time, also referred to as the mean first-passage time (MFPT), which
is useful in the study of transport-limited
reactions~\cite{YuLi02,LoBeMoVo08}, target
search~\cite{BeCoMoSuVo05,Sh06} and other physical problems.

An important question in the study of trapping is how the MFPT
scales with the size of the system. There are some well-known
results providing answers to the corresponding questions in the
cases of some graphs with simple topology, including regular
lattices~\cite{Mo69}, Sierpinski
fractals~\cite{KaBa02PRE,KaBa02IJBC}, T-fractal~\cite{Ag08}, and so
on. However, these graphs are not suitable to describe real
systems~\cite{AlBa02} encountered in everyday experience, most of
which are scale-free~\cite{BaAl99} and small-world~\cite{WaSt98}
that have been shown to influence profoundly various dynamical
processes running on networks~\cite{Ne03,DoGoMe08}. Thus, it is
natural and interesting to explore the trapping problem on networks
with general structure embedded in real life. Although a lot of
activities have been devoted to studying random walks on complex
networks\cite{NoRi04,SoRebe05,Bobe05,BaCaPa08,CaAb08}, work about
trapping problem on scale-free small-world graphs is much
less~\cite{KiCaHaAr08}.

In the paper, we investigate the trapping problem on the Apolloian
network~\cite{AnHeAnSi05,DoMa05} with scale-free and small-world
properties. We focus on a specific aspect of random walks in the
presence of a single trap situated at a given node with the largest
degree (hub node). We obtain an exact analytical solution for the
MFPT and the dependence of this primary quantity on the system size.
We show that the Apolloian network is a preferred architecture that
minimizes the increase of MFPT with network size, compared with
regular networks and fractals.

\section{Introduction to the Apollonian network}

\begin{figure}
\begin{center}
\includegraphics[width=.32\linewidth,trim=20 50 70 35]{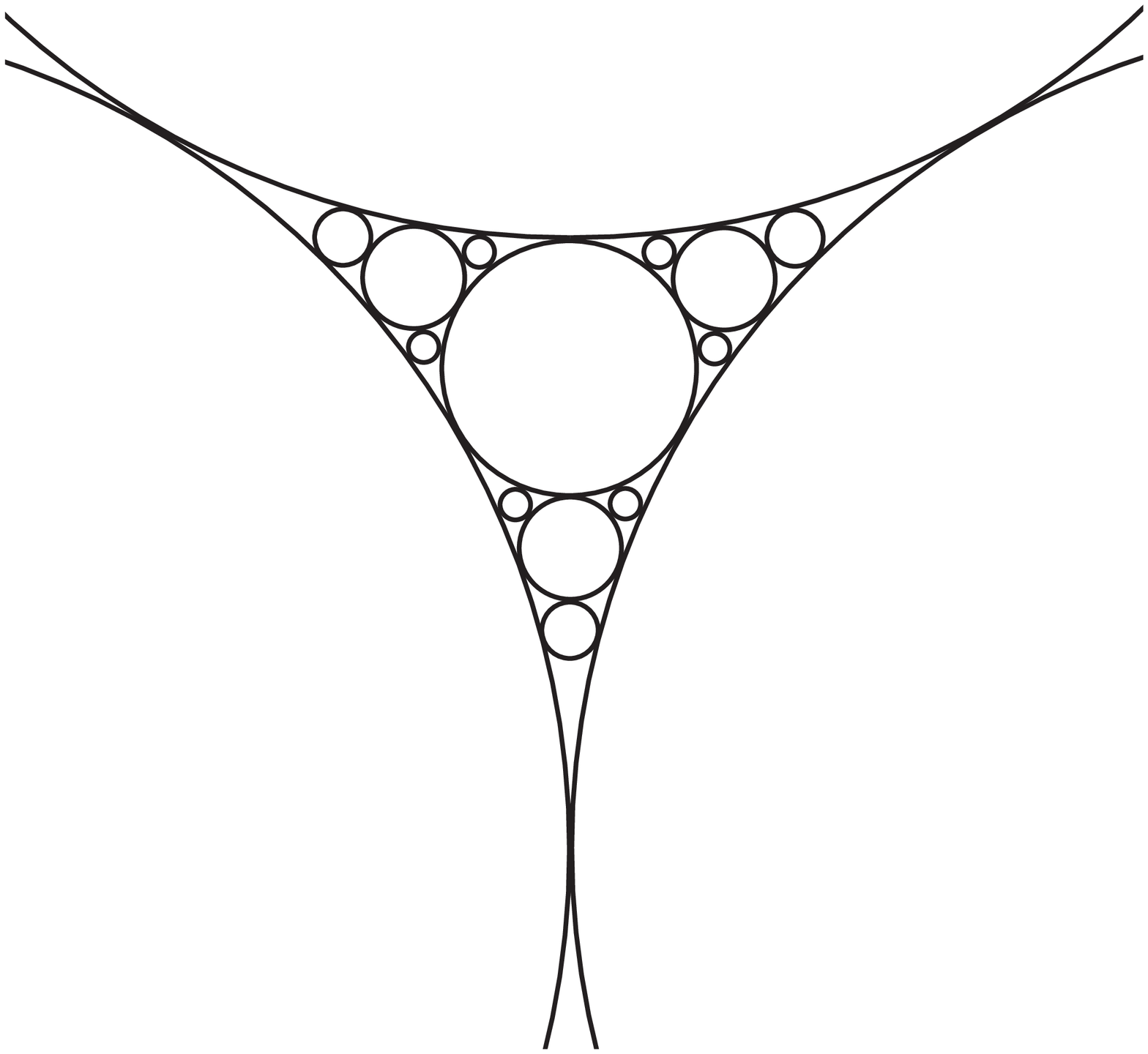}
\includegraphics[width=.45\linewidth,trim=70 50 70 35]{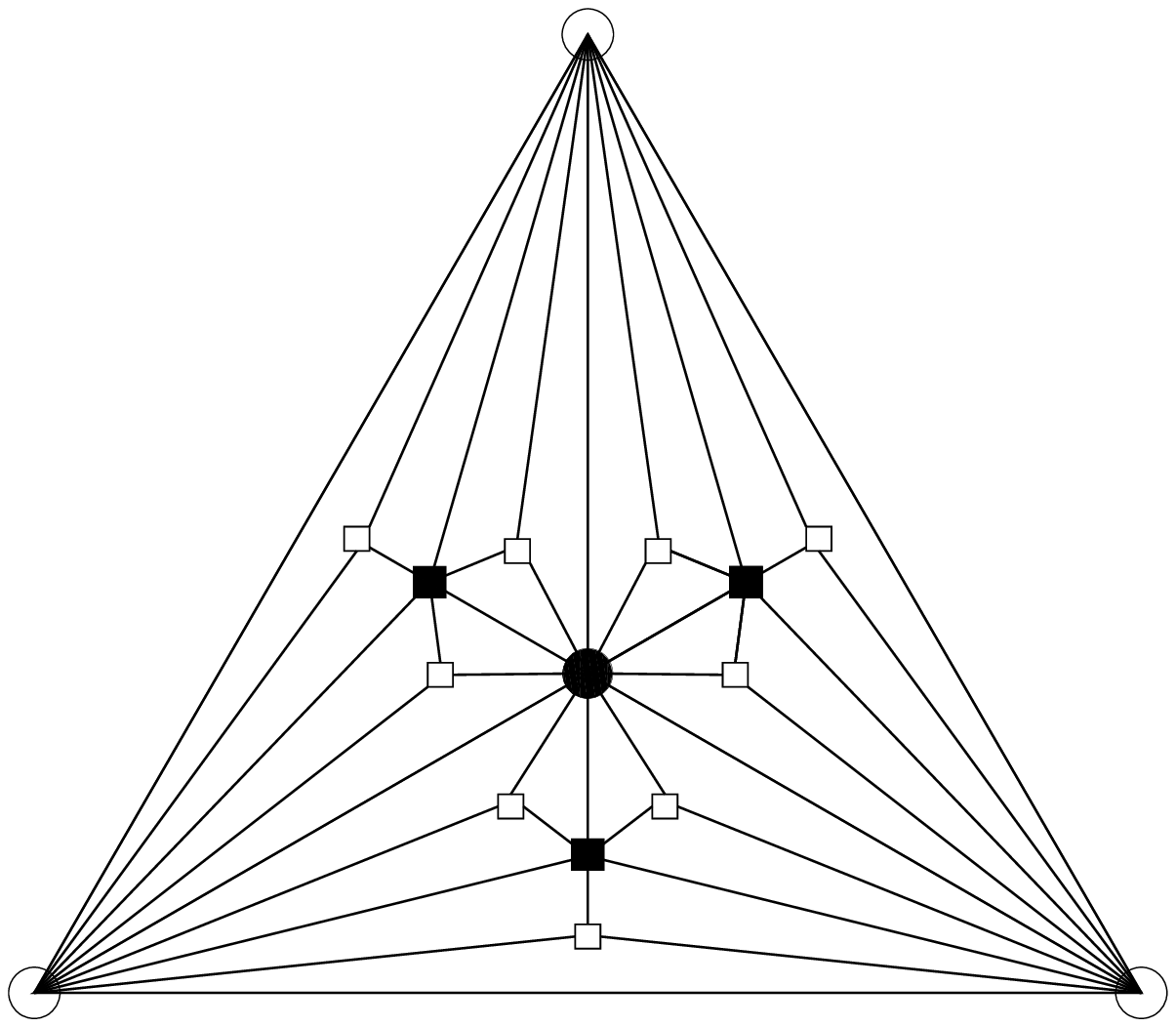}
\end{center}
\caption{(Left panel) The first three generations of the first
Apollonian packing of disks. (Right panel) An Apollonian network
corresponding to the packing shown in the left panel.}
\label{FigApollo1}
\end{figure}

We first introduce the Apollonian packing~\cite{MaHe91}, from which
the Apollonian network is derived. There are two commonly used
Apollonian packings that differ mainly in initial configurations.
The first packing is constructed by starting with three mutually
touching disks, the interstice of which is a curvilinear triangle.
In the first generation a disk is inscribed, touching all the sides
of this curvilinear triangle. For subsequent generations we
indefinitely repeat the packing process for all the new curvilinear
triangles. In the limit of infinite generations, an Apollonian
packing is obtained. The left panel of figure~\ref{FigApollo1}
displays the first three generations of this Apollonian packing.

%%%%%%%%%%%%%%%%%%%%%%%%%%%%%%%%%%%%%%%%%%%%%%%%%%%%%%%%%%
% Figure  1
%%%%%%%%%%%%%%%%%%%%%%%%%%%%%%%%%%%%%%%%%%%%%%%%%%%%%%%%%%
%%%%%%%%%%%%%%%%%%%%%%%%%%%%%%%%%%%%%%%%%%%%%%%%%%%%%%%%%%

\begin{figure}
\begin{center}
\includegraphics[width=.4\linewidth,trim=60 25 60 25]{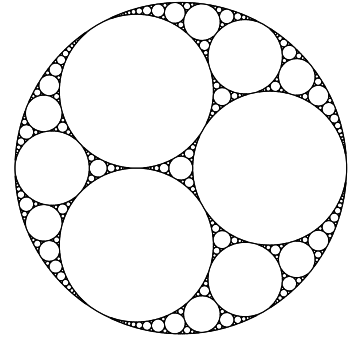}
\caption{The first several generations of the second Apollonian
packing of disks within a circle.} \label{FigApollo02}
\end{center}
\end{figure}

The other frequently used Apollonian packing has a initial
configuration with three mutually touching disks inscribed inside a
circular space. The interstices of the initial disks and circle are
four curvilinear triangles to be filled. Then in each subsequent
generation, we add one disk to each interstice (curvilinear
triangle), so that the added disk touches all the three sides of the
corresponding curvilinear triangle. The first several processes are
shown in figure~\ref{FigApollo02}.

%%%%%%%%%%%%%%%%%%%%%%%%%%%%%%%%%%%%%%%%%%%%%%%%%%%%%%%%%%
% Figure  2
%%%%%%%%%%%%%%%%%%%%%%%%%%%%%%%%%%%%%%%%%%%%%%%%%%%%%%%%%%
%%%%%%%%%%%%%%%%%%%%%%%%%%%%%%%%%%%%%%%%%%%%%%%%%%%%%%%%%%

The above two Apollonian packings can be easily mapped to networks,
usually called as Apollonian networks~\cite{AnHeAnSi05,DoMa05}. The
translation from Apollonian packing construction to Apollonian
network generation is quite straightforward. Each node (vertex,
site) corresponds to a disk, and two nodes are linked to each other
if their corresponding disks are tangent. Note that for the second
Apollonian packing, the initial circle also corresponds to a node.
The right panel of figure~\ref{FigApollo1} shows an example of the
network. Since the resulting networks associated with the two
packings have similar structural features, except for the initial
disks, in this paper we will focus on the network based on the
second Apollonian packing, which is convenient for analytically
deriving the network properties.

%%%%%%%%%%%%%%%%%%%%%%%%%%%%%%%%%%%%%%%%%%%%%%%%%%%%%%%%%%
% Figure  3
%%%%%%%%%%%%%%%%%%%%%%%%%%%%%%%%%%%%%%%%%%%%%%%%%%%%%%%%%%
\begin{figure}
\begin{center}
\includegraphics[width=.45\linewidth,trim=70 50 70 40]{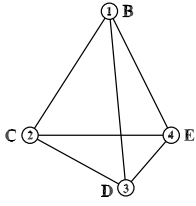}
\end{center}
\caption{The initial construction of Apollonian network.}
\label{FigIni}
\end{figure}
%%%%%%%%%%%%%%%%%%%%%%%%%%%%%%%%%%%%%%%%%%%%%%%%%%%%%%%%%%

\begin{figure}
\begin{center}
\includegraphics[width=0.62\linewidth,trim=100 30 100 30]{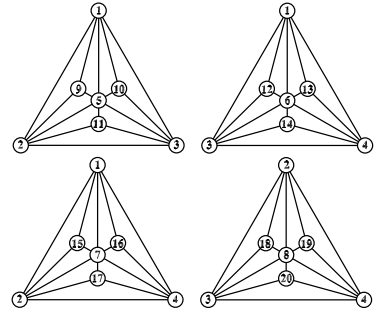} \
\end{center}
\caption[kurzform]{The Apollonian network of generation 2 and the
labeling of its nodes.} \label{network}
\end{figure}

According to the foregoing mapping, one can introduce a general
algorithm~\cite{ZhCoFeRo06,ZhRoZh06} to create the Apollonian
network, denoted by $\mathbb{A}_g$ after $g$ generation evolutions.
For $g=0$, $\mathbb{A}_0$ is a tetrahedron with four faces or
triangles, see figure~\ref{FigIni}. For $g\geq 1$, $\mathbb{A}_g$ is
obtained from $\mathbb{A}_{g-1}$. For each of the existing triangles
of $\mathbb{A}_{g-1}$ that is created at generation $t-1$, a new
node is added and connected to all the three nodes of this triangle.
Figure~\ref{network} illustrates the construction process for the
first two generations of the initial four faces as shown in
figure~\ref{FigIni}.

%%%%%%%%%%%%%%%%%%%%%%%%%%%%%%%%%%%%%%%%%%%%%%%%%%%%%%%%%
% Figure  4
%%%%%%%%%%%%%%%%%%%%%%%%%%%%%%%%%%%%%%%%%%%%%%%%%%%%%%%%%%
%%%%%%%%%%%%%%%%%%%%%%%%%%%%%%%%%%%%%%%%%%%%%%%%%%%%%%%%%%

According to the network construction, one can
see~\cite{DoMa05,ZhRoZh06} that at the $g_i$th ($g_i\geq 1$)
generation, the number of newly introduced nodes is
$L_v(g_i)=4\times 3^{g_i-1}$. From this result, we can easily
compute the total number of nodes $N_g$ in network $\mathbb{A}_g$
(hereafter called the network order):
\begin{equation}\label{Nt}
N_g=\sum_{g_i=0}^{g}L_v(g_i)=2\times3^{g}+2.
\end{equation}

For convenient use in the next section, we distinguish different
nodes of $\mathbb{A}_g$ by labeling them sequentially as follows.
The four nodes (i.e., nodes $B$, $C$, $D$, and $E$ in
figure~\ref{FigIni}) created at initial generation are labeled as 1,
2, 3, and 4, respectively. Then, in each new generation, we only
label the new nodes added at this generation, while keep the labels
of pre-existing nodes unchanged. That is to say, we label new nodes
progressively as $M+1, M+2,\ldots, M+\Delta M$, in which $M$ is the
total number of the pre-existing nodes and $\Delta M$ is the number
of newly-added nodes. Eventually, every node is labeled by a unique
integer, all nodes in $\mathbb{A}_g$ are labeled from 1 to $N_g$,
see figure~\ref{network}.

Let $k_i(g)$ be the degree of a node $i$ at generation $g$, which
entered the network at the generation of $g_i$ ($g_i\geq 0$).
Then~\cite{DoMa05,ZhRoZh06}
\begin{equation}\label{ki}
k_i(g)=3\times2^{g-g_{i}}.
\end{equation}
From equation~(\ref{ki}), one can easily see that at each step the
degree of a node doubles, i.e.,
\begin{equation}\label{ki2}
k_i(g)=2\,k_i(g-1).
\end{equation}

The Apollonian network presents the typical characteristics of
real-life
networks~\cite{AnHeAnSi05,DoMa05,ZhCoFeRo06,ZhRoZh06,ZhChZhFaGuZo08}.
It has a power-law degree distribution $P(k) \sim k^{-\gamma}$ with
the exponent $\gamma=1+\frac{\ln 3}{\ln 2}$. Its average path
length, defined as the mean of shortest distance between all pairs
of nodes, increases logarithmically with network
order~\cite{ZhChZhFaGuZo08}. In the large network order limit, the
average clustering coefficient tends to $0.8284$. Thus, the
Apollonian network exhibits small-world effect~\cite{WaSt98}. In
addition, the network is disassortative~\cite{DoMa05}.
\textcolor{blue} {In view of its structural properties similar to
those of real networks and its intrinsic interest, Apollonian
network is a good substrate network for studying criticality
phenomena and dynamical
processes~\cite{ZhYaWa05,HuXuWuWa06,HaMa06,MoPaFiAn06,LiSiAnHe06,ScGoMoAnHe07,ViAnHeAn07,OlMoLyAnAl09,KaHiBe09}.}
In what follows we will investigate the MFPT for random walks with
an immobile trap on the Apollonian network.

\section{Formulation of the problem}

In this section we formulate the problem of a simple random walk of
a particle on Apollonian network $\mathbb{A}_g$ in the presence of a
trap or a perfect absorber positioned on a given node. To this end,
we first specify $\mathbb{A}_g$ by its adjacency matrix
$\textbf{A}_g$ (hereafter we neglect the subscript) of order
$N_g\times N_g$, which completely describes network $\mathbb{A}_g$.
The element $a_{ij}$ of $\textbf{A}$ is defined as follows:
$a_{ij}=1$ if the pair of node $i$ and node $j$ is connected by a
link (edge, bond), otherwise $a_{ij}=0$. The degree $k_i(g)$ of node
$i$ is $k_i(g)=\sum_{j=1}^{N_g}a_{ij}$, and the diagonal degree
matrix $\mathbf{Z}$ is given by $\textbf{Z}={\rm diag} (k_1(g),
k_2(g),\cdots, k_i(g), \cdots, k_{N_{g}}(g))$. Then, the normalized
Laplacian matrix of $\mathbb{A}_g$ is given by
$\textbf{L}=\textbf{I}-\textbf{Z}^{-1}\textbf{A}$ and its entry is
$l_{ij}=1-\frac{a_{ij}}{k_i(g)}$, where $\textbf{I}$ is an identity
matrix.

Before proceeding further, let us introduce the so-called simple
random walk~\cite{HaBe87,MeKl04,BuCa05} on network $\mathbb{A}_g$.
At each time step (taken to be unity), the walker moves from its
current location to any of its nearest neighbors with equal
probability. According to this rule, at time $t$, a walker at a node
$i$ will hop to one of its $k_i(g)$ neighbors, say $u$, with the
transition probability $a_{iu}/k_i(g)$. Suppose that the walker
starts from node $i$ at $t=0$, then the jumping probability $P_{ij}$
of going from $i$ to $j$ at time $t$ is governed by the following
master equation:
\begin{equation}\label{master}
 P_{ij}(t+1)=\sum_{w=1}^{N_g} \frac{ a_{wj}}{ k_w(g)} \, P_{iw}(t).
\end{equation}

We next focus a random walk on $\mathbb{A}_g$ with a trap. We locate
the trap at node 1, represented as $i_T$. It should be pointed out
that thanks to the symmetry, the trap can be also situated at node
2, 3, or 4, which has no influence on MFPT. The particular selection
for the trap location allows one to calculate analytically the MFPT,
which will be addressed in detail in the next section. At each time
step, the walker, starting from any node except the trap $i_T$,
jumps to any of its nearest neighbors with equal probability.
%It is not difficult to verify that the Markov chain representing such a random walk is ergodic. That is to say, the walker will be absorbed by the trap node, regardless of its origin~\cite{Bobe05}.

Let $T_i$ be mean transmit time (first-passage time, trapping time,
or the mean time to absorption) for a walker, starting from node
$i$, to first arrive at the trap $i_T$. Thus, $T_{i_T}=0$. Then, the
set of these interesting quantities follows the recurrence
equation~\cite{KeSn76}
\begin{equation}\label{MFPT1}
 T_i=\sum_{j=2}^{N_g} \frac{ a_{ij}}{ k_i(g)}\,T_j+1,
\end{equation}
where $i\neq i_T$. Equation~(\ref{MFPT1}) may be also rewritten in
matrix notation as
\begin{equation}\label{MFPT2}
 \textbf{T}=\mathbf{\Delta}^{-1}\,\textbf{e},
\end{equation}
where $\textbf{T}=(T_2,T_3,\cdots,T_{N})^\top$ and
$\textbf{e}=(1,1,\cdots,1)^\top$ are two ($N_g-1$)-dimensional
vectors, and $\mathbf{\Delta}^{-1}$ is the fundamental matrix of the
Markov chain representing such unbiased random walk. In fact,
$\mathbf{\Delta}$ is a sub-matrix of the normalized discrete
Laplacian matrix $\mathbf{L}$ whose first row and column,
corresponding to the trap node, have been removed.

Then, the mean first-passage time, or the average of the mean time
to absorption, $\langle T \rangle_g$, which is the average of $T_i$
over all nodes distributed uniformly over nodes in $\mathbb{A}_g$
other than the trap, is given by
\begin{equation}\label{MFPT5}
 \langle T
\rangle_g=\frac{1}{N_g-1}\sum_{i=2}^{N_g}
T_i=\frac{1}{N_g-1}\sum_{i=2}^{N_g}\sum_{j=2}^{N_g}{(\Delta^{-1})_{ij}}.
\end{equation}

Equation~(\ref{MFPT5}) can be easily explained by looking at the
random walk from the perspective of a Markov chain. Actually, the
entry $(\Delta^{-1})_{ij}$ of the fundamental matrix
$\mathbf{\Delta}^{-1}$ of the Markov process expresses the average
number of times that a walker starting at node $i$ will be at node
$j$, and the row sum $\sum_{j=2}^{N_g}{(\Delta^{-1})_{ij}}$ is
exactly $T_i$, the total times a particle starting at node $i$ will
traverse all other nodes before being absorbed by the trap.

Equation~(\ref{MFPT5}) shows that the problem of determining
$\langle T \rangle_g$ is reduced to finding the sum of all elements
of the fundamental matrix $\mathbf{\Delta}^{-1}$ of order
$(N_g-1)\times (N_g-1)$, \textcolor{blue} {which can be obtained by
utilizing a standard software package, Mathematica 5.0.}
Irrespective of the seemingly compact expression of
equation~(\ref{MFPT5}), since $N_g$ increases exponentially with
$g$, for large $g$, it becomes difficult to obtain $\langle T
\rangle_g$ through direct calculation from this equation, because of
the limitations of time and computer memory. Therefore, one can
compute directly the MFPT only for the first generations, see
table~\ref{tab:AMTA1}. However, the particular construction of the
Apollonian network allows one to calculate analytically MFPT to
obtain a rigorous solution. Details of derivation will be provided
below.

%%%%%%%%%%%%%%%%%%%%%%%%%%%%%%%%%%%%%%%%%%%%%%%%
%% Table 1
%%%%%%%%%%%%%%%%%%%%%%%%%%%%%%%%%%%%%%%%%%%%%%%%%
\begin{table}
\caption{The MFPT obtained by direct calculation from
equation~(\ref{MFPT5}) \textcolor{blue} {by using a standard
software package, Mathematica 5.0.} Since for large networks, the
computation of the MFPT from equation~(\ref{MFPT5}) is prohibitively
time and memory consuming, we calculate the MFPT for the first
several generations.} \label{tab:AMTA1}
\begin{center}
\begin{tabular}{ccccc}
\hline \hline
\quad\quad $g$\quad\quad\quad & \quad\quad\  $N_g$ \quad\quad\quad  &\quad\quad\quad $\langle T\rangle_g$ \quad\quad\quad \\
\hline
0  & 4 & $9/3$  \\
1  & 8 &  $182/(5 \times 7)$ \\ %$\frac{182}{5 \times 7}$
2  & 20 & $861/(5 \times 19)$  \\%\frac{861}{5\times19}
3 & 56 & $109854/(125 \times 55)$ \\%\frac{109854}{125 \times 55}
4  & 164 & $2895129/(625 \times 163)$  \\%\frac{2895129}{625 \times 163}
5  & 488 & $77327622/(3125 \times 487)$  \\%\frac{77327622}{3125 \times 487}
6  & 1460 & $415448109/(3125 \times 1459)$   \\%\frac{415448109}{3125 \times 1459}
\hline \hline
\end{tabular}
\end{center}
\end{table}
%%%%%%%%%%%%%%%%%%%%%%%%%%%%%%%%%%%%%%%%%%%%%%%%

\section{Exact solution for mean first-passage time}

Before deriving the general formula for MFPT, $\langle T \rangle_g$,
we first establish the scaling relation dominating the evolution of
$T_i^g$ with generation $g$, where $T_i^g$ is the trapping time for
a walk originating at node $i$ on the $g$th generation of Apollonian
network.

\subsection{Evolution scaling for trapping time}

%%%%%%%%%%%%%%%%%%%%%%%%%%%%%%%%%%%%%%%%%%%%%%%%
%% Table 2
%%%%%%%%%%%%%%%%%%%%%%%%%%%%%%%%%%%%%%%%%%%%%%%%%
\begin{table*}
\caption{Mean time to absorption $T_i^g$ for a random walker
starting from node $i$ on the Apollonian network for various $g$.
Notice that owing to the obvious symmetry, nodes in a parenthesis
are equivalent, since they have the same trapping time. All the
values are calculated straightforwardly from
equation~(\ref{MFPT5}).} \label{tab:AMTA2}
\begin{center}
\begin{tabular}{l|cccccccccc}
\hline \hline  $g\backslash i$  & \quad\quad(2,3,4)\quad\quad\quad & \quad\quad(5,6,7)\quad \quad\quad & \quad \quad (8)\quad\quad \quad &(9,10,12,13,15,16) &\quad(11,14,17)\quad &\quad (18,19,20)\quad \quad \\
\hline
0 & $3$       \\
1 & $27/5$ & $23/5$ & $32/5$ \\
2 & $243/25$&    $207/25$&           $288/25$ &      $7$ &       $256/25$ &      $283/25$\\
3 & $2187/125$&  $1863/125$&         $2592/125$ &    $63/5$ &    $2304/125$ &    $2547/125$  \\
4 & $19683/625$ &  $16767/625$ & $23328/625$ &
            $567/25$ &  $20736/625$ &   $22923/625$  \\
5 & $177147/3125$&      $150903/3125$& $209952/3125$ &
            $5103/125$& $186624/3125$& $206307/3125$  \\
6 & $1594323/15625$&    $1358127/15625$& $1889568/15625$&
            $45927/625$&$1679616/15625$&$1856763/15625$  \\
\hline \hline
\end{tabular}
\end{center}
\end{table*}
%%%%%%%%%%%%%%%%%%%%%%%%%%%%%%%%%%%%%%%%%%%%%%%%

We begin by recording the numerical values of $T_i^g$. Obviously,
for all $g \geq 0$, $T_1^g=0$; for $g = 0$, it is a trivial case, we
have $T_2^0=T_3^0=T_4^0=3$. For $g \geq 1$, the values of $T_i^g$
can be obtained straightforwardly via equation~(\ref{MFPT5}). Table
\ref{tab:AMTA2} lists the numerical values of $T_i^g$ for some nodes
up to $g=6$. The numerical values listed in table \ref{tab:AMTA2}
show that for a given node $i$ we have
$T_i^{g+1}=\frac{9}{5}\,T_i^g$. That is to say, upon growth of the
Apollonian network from $g$ to generation $g+1$, the mean time to
first reach the trap increases by a factor $\frac{9}{5}$. This is a
basic character of random walks on the Apollonian network, which can
be established from the arguments below~\cite{HaBe87,Bobe05}.

\begin{figure}
\begin{center}
\includegraphics[width=.5\linewidth,trim=60 50 60 40]{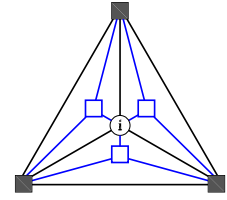}
\end{center}
\caption{Growth of trapping time in going from $\mathbb{A}_{g}$ to
$\mathbb{A}_{g+1}$. Node $i \in \mathbb{A}_{g}$ has $k_i$ neighbor
nodes in generation $g$ ($\blacksquare$) and $k_i$ new neighbor
nodes in generation $g+1$ ($\square$). A new node has a degree of 3,
and is linked to $i$ and its two old neighbor nodes.} \label{Trap}
\end{figure}

Consider an arbitrary node $i$ in the Apollonian network
$\mathbb{A}_g$ after $g$ generation evolution. From
equation~(\ref{ki}), we know that upon growth of the Apollonian
network to generation $g+1$, the degree $k_i$ of node $i$ doubles.
\textcolor{blue} {Thus, at generation $g+1$, node $i$ has $k_i$ old
neighbors (born at generation $g$ or earlier) and $k_i$ new
neighbors (generated at generation $g+1$).}  Let the mean transmit
time for going from node $i$ to any of its $k_i$ old neighbors be
$X$; and let the mean transmit time for going from any of its $k_i$
new neighbors to one of the $k_i$ old neighbors be $Y$. Then we can
establish the following underlying backward equations (see
figure~\ref{Trap})
\begin{equation}\label{MFPT6}
X=\frac{1}{2}+\frac{1}{2}(1+Y),\quad \quad \quad
Y=\frac{2}{3}+\frac{1}{3}(1+X),
\end{equation}
which leads to $X=\frac{9}{5}$. That is to say, the passage time
from any node $i$ ($i \in \mathbb{A}_{g}$) to any node $j$ ($j\in
\mathbb{A}_{g}$) increases by a factor of $\frac{9}{5}$, upon the
network growth from generation $g$ to generation $g+1$. Thus, we
have $T_i^{g+1}=\frac{9}{5}\,T_i^g$, which will be useful for
deriving the formula for the mean first-passage time in the
following text.

%%%%%%%%%%%%%%%%%%%%%%%%%%%%%%%%%%%%%%%%%%%%%%%%%%%%%%%%%%
% Figure  5
%%%%%%%%%%%%%%%%%%%%%%%%%%%%%%%%%%%%%%%%%%%%%%%%%%%%%%%%%%
%%%%%%%%%%%%%%%%%%%%%%%%%%%%%%%%%%%%%%%%%%%%%%%%%%%%%%%%%%

\subsection{Formula for the mean first-passage time}

Having obtained the scaling of mean transmit time for old nodes, we
now determine the mean first-passage time, with an aim to derive an
exact solution. We represent the set of nodes in $\mathbb{A}_{g}$ as
$\Omega_g$, and denote the set of nodes created at generation $g$ by
$\overline{\Omega}_g$. Thus we have
$\Omega_g=\overline{\Omega}_g\cup\Omega_{g-1}$. For the convenience
of computation, we define the following quantities for $m \leq g$:
\begin{equation}\label{MFPT7}
 T_{m,{\rm total}}^g=\sum_{i\in \Omega_m} T_i^g,
\end{equation}
and
\begin{equation}\label{MFPT8}
 \overline{T}_{m,{\rm total}}^g=\sum_{i\in \overline{\Omega}_m}
 T_i^g.
\end{equation}
Then, we have
\begin{equation}\label{MFPT9}
 T_{g, {\rm total}}^g=T_{g-1,{\rm total}}^g+\overline{T}_{g, {\rm total}}^g.
\end{equation}
Next we will explicitly determine the quantity $T_{g,{\rm
total}}^g$. To this end, we should firstly determine
$\overline{T}_{g, {\rm total}}^g$.

We examine the mean time to absorption for the first several
generations of Apollonian network. In the case of $g=1$, by the very
construction of Apollonian network, it follows that
\begin{eqnarray}
T^{1}_5&=&(1+T^{1}_1)/3+(1+T^{1}_2)/3+(1+T^{1}_3)/3,\nonumber\\
T^{1}_6&=&(1+T^{1}_1)/3+(1+T^{1}_3)/3+(1+T^{1}_4)/3,\nonumber\\
T^{1}_7&=&(1+T^{1}_1)/3+(1+T^{1}_2)/3+(1+T^{1}_4)/3,\nonumber\\
T^{1}_8&=&(1+T^{1}_2)/3+(1+T^{1}_3)/3+(1+T^{1}_4)/3.\nonumber
\end{eqnarray}
Thus,
\begin{eqnarray}\label{MFPT10}
\overline{T}_{1,{\rm total}}^1&=&\sum_{i\in
\overline{\Omega}_1}T^{1}_i=T^{1}_5+T^{1}_6+T^{1}_7+T^{1}_8\nonumber\\
&=&4+(T^{1}_1+ T^{1}_2+ T^{1}_3)=4+\overline{T}_{0,{\rm total}}^1\,.
\end{eqnarray}
Similarly, for $g=2$ case, it is easy to obtain
\begin{eqnarray}\label{MFPT11}
\overline{T}_{2,{\rm total}}^2&=&\sum_{i\in
\overline{\Omega}_2}T^{2}_i=\sum_{i=9}^{20}T^{2}_i\nonumber\\
&=&4\times 3+2\,\overline{T}_{0,{\rm total}}^2+\overline{T}_{1,{\rm
total}}^2\,.
\end{eqnarray}
Proceeding analogously, it is not difficult to derive that
\begin{eqnarray}\label{MFPT12}
\overline{T}_{g,{\rm total}}^g=4\times 3^{g-1}&+&\overline{T}^{g}_{g-1,{\rm total}}+2\,\overline{T}^{g}_{g-2,{\rm total}}+\dots\nonumber\\
&+&2^{g-2}\,\overline{T}^{g}_{1,{\rm
total}}+2^{g-1}\,\overline{T}^{g}_{0,{\rm total}},
\end{eqnarray}
and
\begin{eqnarray}\label{MFPT13}
\overline{T}_{g+1,{\rm total}}^{g+1}=4\times 3^{g}&+&\overline{T}^{g+1}_{g,{\rm total}}+2\,\overline{T}^{g+1}_{g-1,{\rm total}}+\dots\nonumber\\
&+&2^{g-1}\,\overline{T}^{g+1}_{1,{\rm
total}}+2^{g}\,\overline{T}^{g+1}_{0,{\rm total}},
\end{eqnarray}
where $4\times 3^{g-1}$ and $4\times 3^{g}$ are actually the numbers
of nodes generated at generations $g$ and $g+1$, respectively.
Equation~(\ref{MFPT13}) minus equation~(\ref{MFPT12}) times
$\frac{18}{5}$ and making use of the relation
$T_i^{g+1}=\frac{9}{5}\,T_i^g$, one gets
\begin{equation}\label{MFPT15}
\overline{T}^{g+1}_{g+1,{\rm
total}}=\frac{27}{5}\,\overline{T}^{g}_{g,{\rm
total}}-\frac{4}{5}\times 3^g.
\end{equation}
Using $\overline{T}_{1,{\rm total}}^1=\frac{101}{5}$,
equation~(\ref{MFPT15}) is solved inductively
\begin{equation}\label{MFPT16}
\overline{T}^{g}_{g,{\rm
total}}=\frac{32}{9}\left(\frac{27}{5}\right)^g+3^{g-1}\,.
\end{equation}
Substituting equation~(\ref{MFPT16}) for $\overline{T}^{g}_{g,{\rm
total}}$ into equation~(\ref{MFPT9}) and using $T_{g-1,{\rm
total}}^g=\frac{9}{5}\,T_{g-1,{\rm total}}^{g-1}$, we have
\begin{equation}\label{MFPT17}
T_{g,{\rm total}}^g=\frac{9}{5}\,T_{g-1,{\rm
total}}^{g-1}+\frac{32}{9}\left(\frac{27}{5}\right)^g+3^{g-1}\,.
\end{equation}
Considering the initial condition $T_{1,{\rm
total}}^1=\frac{182}{5}$, equation~(\ref{MFPT17}) is resolved by
induction to yield
\begin{equation}\label{MFPT18}
T_{g,{\rm total}}^g=\frac{3^g}{6 \times 5^g}(17 \times 3^g + 5^{g
         + 1} + 32 \times 9^g) \,.
\end{equation}
Plugging the last expression into equation~(\ref{MFPT5}), we arrive
at the accurate formula for the average of the mean time to
absorption at the trap located at node 1 on the $g$th of Apollonian
network:
\begin{equation}\label{MFPT19}
 \langle T\rangle_g =\frac{3^g(17 \times 3^g + 5^{g
         + 1} + 32 \times 9^g)}{6 \times 5^g(2\times 3^g+1)}\,.
\end{equation}
We have checked our analytic formula against numerical values quoted
in Table~\ref{tab:AMTA1}. For the range of $0 \leq g \leq 6$, the
values obtained from equation~(\ref{MFPT19}) completely agree with
those numerical results on the basis of the direct calculation
through equation~(\ref{MFPT5}). This agreement serves as an
independent test of our theoretical formula.

From equation~(\ref{Nt}), we have
$g=\log_3\big(\frac{N_g}{2}-1\big)$. Hence, for large network (i.e.,
$N_g\rightarrow \infty$), we obtain
\begin{equation}\label{MFPT21}
\langle T \rangle_g \sim \left(\frac{9}{5}\right)^g=
\left(\frac{N_g}{2}-1\right)^{2-\frac{\ln5}{\ln3}}\sim
N_g^{(2-\ln5/\ln3)},
\end{equation}
where the exponent $2-\frac{\ln5}{\ln 3} \approx 0.535 <1$. Thus, in
the large limit of network order $N_g$, the MFPT increases
algebraically with increasing order of the network.

\begin{figure}
\begin{center}
\includegraphics[width=.3\linewidth,trim=100 30 100 40]{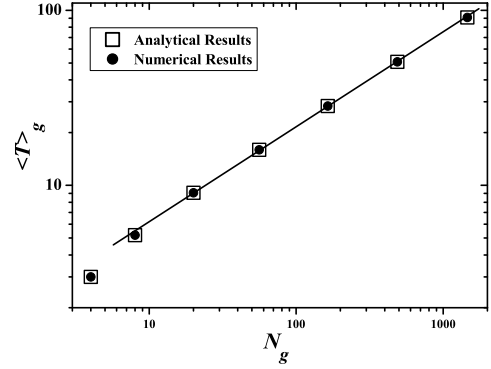}
\end{center}
\caption[kurzform]{\label{MTT} Mean first-passage time $\langle T
\rangle_g$ versus network order $N_{g}$ on a log-log scale. The
solid line is a guide to the eye.}
\end{figure}

\textcolor{blue} {Generally, finite-size effect plays a crucial role
in critical phenomena and dynamical processes on
networks~\cite{HoHaPa07}. In order to study the finite-size effect
in the scaling behavior of MFPT for the Apollonian network, we plot
a figure to show the MFPT with order for small networks, see
figure~\ref{MTT}. Surprisingly, it is observed from figure~\ref{MTT}
that the finite-size effect has little influence on the scaling.}

%%%%%%%%%%%%%%%%%%%%%%%%%%%%%%%%%%%%%%%%%%%%%%%%%%%%%%%%%
% Figure  6
%%%%%%%%%%%%%%%%%%%%%%%%%%%%%%%%%%%%%%%%%%%%%%%%%%%%%%%%%%
%%%%%%%%%%%%%%%%%%%%%%%%%%%%%%%%%%%%%%%%%%%%%%%%%%%%%%%%%%

Previous studies showed that for regular lattices~\cite{Mo69},
Sierpinski fractals~\cite{KaBa02PRE,KaBa02IJBC}, and
T-graph~\cite{Ag08} with large order $N$, their MFPT $ \langle T
\rangle $ behaves as $ \langle T \rangle \sim N^{\alpha}$ with
$\alpha >1$. Even in a complete graph of $N$ nodes, $K_N$, the MFPT
$ \langle T \rangle $ grows linearly with $N$ as $ \langle T \rangle
=N-1$. Actually, linear scaling of the MFPT with $N$ is the best
that has been reported so far. From equation~(\ref{MFPT21}), one can
see that the MFPT $ \langle T \rangle $ of the Apollonian network
increases as a fractional power of network order $N$, which implies
that Apollonian network has a faster transmit time than any other
analytically soluble media. In other words, the Apollonian network
has the best structure for fast diffusion one can see heretofore.

\textcolor{blue} {We argue that the heterogeneous topology of the
Apollonian network may be responsible for the high efficiency for
transport. In the Apollonian network, there are some nodes with high
degree, which are linked to one another and to most other nodes in
the network. These `large' nodes can be easily visited by a walker
starting from an arbitrary location.  Since `large' nodes, including
the trap node, are connected to one another, so the walker can find
the trap in a very short time.}

\section{Conclusions}

In summary, we have studied trapping problem on the Apollonian
network exhibiting remarkable features (scale-free behavior and
small-world effects) of a variety of real networks. We have obtained
an analytical closed-form solution for the MFPT for random walks
with a trap located at a hub node, which is consistent with the
numerical computation. The rigorous result indicates that the MFPT
on the Apollonian network shows a very distinct behavior, which is
compared with those results previously reported for regular
lattices, Sierpinski fractals, T-graph, and complete graph. We have
shown that Apollonian network is the most efficient network for
transport by diffusion, in contrast to other studied structure. We
hope that the current study may cast some light on trapping problem
in some real networks, that present similar topologies as the
Apollonian network.

\section{Acknowledgment}
This research was supported by the National Basic Research Program
of China under grant No. 2007CB310806, the National Natural Science
Foundation of China under Grant Nos. 60704044, 60873040 and
60873070, Shanghai Leading Academic Discipline Project No. B114, and
the Program for New Century Excellent Talents in University of China
(NCET-06-0376).

\end{document}